# A Magnetically Switchable Bifocal Metasurface

Alberto Santonocito[1,2], Barbara Patrizi[1], Alessio Gabbani[2], Francesco Pineider*[2], Guido Toci*[1]

[1]Istituto Nazionale di Ottica (INO) Consiglio Nazionale delle Ricerche (CNR), Via Madonna del Piano 10, 50019 Sesto Fiorentino (FI), Italy

[2]Dipartimento di Chimica e Chimica Industriale, Università di Pisa, Via Moruzzi 13, 56124 Pisa (PI), Italy

*e-mail: guido.toci@ino.cnr.it; francesco.pineider@unipi.it

**Abstract:** Tunable flat optics are essential for advancing compact photonic devices. Here we show a numerical study of a reflective magneto-optical metasurface with a dynamically tunable focal length. The structure comprises bismuth iron garnet nanodisks in a Gires-Tournois resonator configuration. The magneto-optical properties of the garnet modulate the reflected phase response via an external magnetic field, allowing focusing at different focal lengths. Full-wave simulations demonstrate that the metasurface exhibits distinct focusing characteristics depending on the applied magnetic field direction for a fixed right circularly polarized incident wave at 1.550 μm. Specifically, switching the external field from +0.2 T to -0.2 T changes the focal length by a factor of approximately two (from 7.16 mm to 13.76 mm). These findings demonstrate that magneto-optical metasurfaces offer a flexible, viable approach for non-mechanical, tunable focusing in compact reflective optical components.

## 1. Introduction

Metasurfaces have fundamentally changed the landscape of modern photonics by enabling the precise manipulation of optical wavefronts through subwavelength-scale engineering. These planar devices provide compact, multifunctional alternatives to conventional refractive optics, facilitating advances in diverse fields such as LIDAR[1], holographic projection[2], and autonomous vision systems[3]. However, the vast majority of realized metasurfaces are static, featuring an optical function permanently fixed at the design stage. To meet the demands of next-generation adaptive optical systems, research efforts are increasingly focused on developing reconfigurable metasurfaces capable of dynamic, on-demand modulation[4–6].

While various tuning mechanisms have been explored, including mechanical actuation[7], liquid crystal reorientation[8], and phase-change materials (PCMs)[9], significant challenges remain. For instance, while PCMs offer large refractive index contrasts, they often suffer from high optical absorption in their crystalline states, limiting their efficiency in the visible and near-infrared regimes[10]. Consequently, there is a distinct need for tuning mechanisms that offer low optical loss, high speed, and robust reversibility.

Magneto-optical (MO) effects present a compelling solution. Tuning via external magnetic fields is non-invasive and allows for rapid, limitless switching cycles[11,12]. Although the MO response in bulk materials is typically weak, nanostructured metasurfaces can significantly enhance these effects through electromagnetic resonance[4,13–17].

Recent advances have demonstrated a growing interest in exploiting these phenomena for integrated photonics. Notably, Christofi et al.[18] designed an all-dielectric MO metasurface composed of periodic bismuth iron garnet (BIG) nanodisks embedded into a low-index nonmagnetic matrix. Their work demonstrated that the interference between Mie and lattice resonances produces an electromagnetically induced transparency (EIT)-like effect. This mechanism greatly enhances Faraday rotation while maintaining minimal optical losses, validating the potential of rare-earth iron garnets for high-performance nanophotonics.

Building upon these foundational material studies, we propose and demonstrate a magnetically tunable metasurface designed for the telecommunications C-band ($\lambda$ = 1.550 μm). Unlike transmissive designs, our device employs a Gires-Tournois (GT) resonator architecture. The structure consists of BIG cylindrical resonators situated atop a silicon dioxide ($SiO_2$) spacer layer, which is backed by a high-reflectivity mirror. This configuration effectively folds the optical path, trapping light within the active layer to enhance the magneto-optical phase accumulation.

In this work, we show that this reflective architecture allows for dramatic tuning of the focal position solely by modulating the external magnetic field. We demonstrate that switching the magnetic field direction from +0.2 T to -0.2 T shifts the focal length from $f(+H)$ = 7160.8 μm to $f(-H)$ = 13758.8 μm.

Besides, we have devised an original design process (reported in Methods) in order to define the distribution of resonators across the surface, tailored to obtain a large variation of the focal length of the metasurface, along with a fairly high focusing efficiency, for the two configurations.

This approach establishes a robust framework for developing varifocal metasurfaces that are compact, efficient, and magnetically reconfigurable.

## 2. Results and Discussion

### 2.1 Unit Cell Characterization and Magnetic Tuning

In this section, we present and analyze the electromagnetic response of the proposed unit cell, with particular emphasis on its tunability under an external magnetic bias. The objective of this analysis is to elucidate how magneto-optical effects influence both the amplitude and phase of the reflected wave, key parameters for the realization of a magnetically reconfigurable metasurface. Building on this tunability, we design a reflective metasurface capable of focusing incident light at two distinct focal positions. The focal point can be dynamically switched by reversing the direction of the applied external magnetic field (+*H* and -*H*), without altering either the physical geometry of the structure or the incident polarization.

To establish the foundation of our design, we first characterized the optical response, specifically reflectance and phase delay as a function of pillar diameter, of individual nanostructures. We examined four distinct periodicities to validate our methodology (see Methods). Figure 1 illustrates the unit cell behavior for a representative period of $P$ = 1.000 μm, displaying both the phase delay and reflectance as a function of pillar diameter. It is important to note that, due to the magneto-optical nature of the BIG material, the physics of this system dictates that reversing the magnetic field direction (+H to -H) is optically equivalent to switching the handedness of the incident polarization (from RCP to LCP) while maintaining a fixed field.

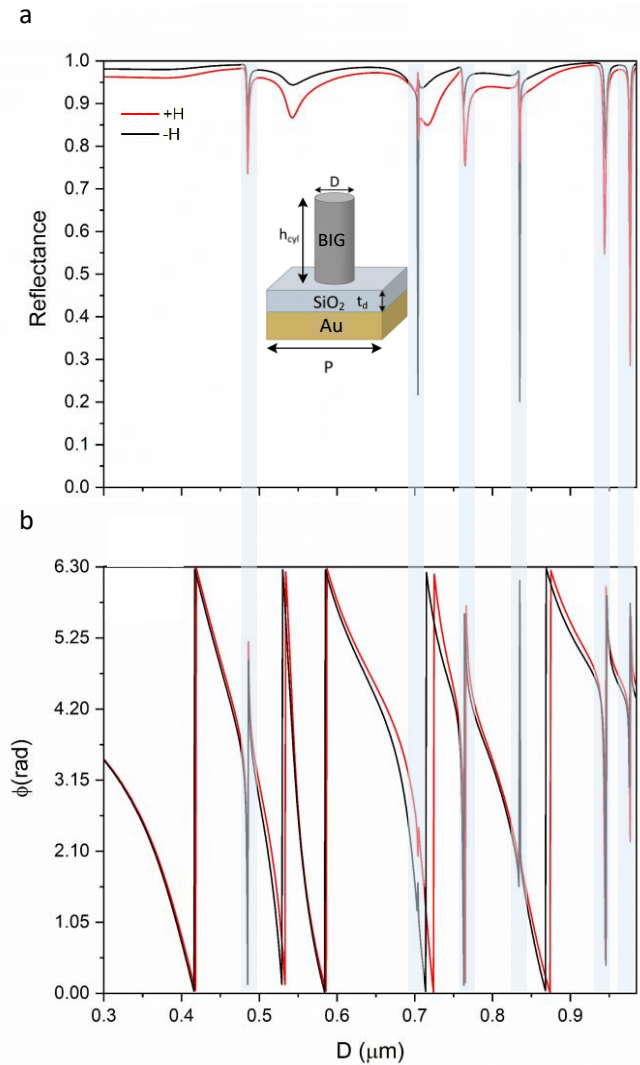

**Figure 1: Optical response of the BIG nanostructures.** (a) Reflectance and (b) phase delay as a function of BIG pillars diameter for incident RCP light in the presence of the magnetic field *+H* (red) and *-H* (black). The pale blue stripes highlight the regions where significant change of both reflectance and phase are localized. In the inset in (a) is reported the unit cell of the system.

## 2.2  Optimization of the Bifocal Metasurface

Building on this unit cell characterization, we tried to identify the optimal focusing configurations. We performed a comprehensive optimization scan, calculating the Figure of Merit (FoM), defined in Methods, Eq. (3), for every combination of focal lengths in the range 2000 μm - 15000 μm. In synthesis, the optimization proceeds as it follows: for each pair of target focal lengths, we calculated the desired phase shift for both focal lengths (Eq. (2) in Methods). Then for each nanopillar position on the metasurface, we search for the individual pillar diameter that minimizes the local phase error for both focal lengths (with $\pm H$) and has high reflection; by this approach, we are able to calculate the best candidate distributions of pillar diameters. Then the focusing efficiency of the metalenses (with $\pm H$) is calculated (as the Strehl ratio) using a full wave propagation method (angular spectrum method). We then finally calculate a Figure of Merit (FoM), based on the harmonic mean of the Strehl ratios for the two focal lengths, which that captures the joint focusing performance of the metalens under $\pm H$ (i.e. for the two focal lengths). This analysis was carried out across different unit cell periods ($P$ = 0.900 μm, 1.000 μm, 1.163 μm, and 1.300 μm) to understand the impact of periodicity on the performance.

The results of this exhaustive search are visualized in Figure 2, which presents the efficiency maps for the considered periods.

The colour gradient represents the magnitude of the calculated FoM, where bright yellow regions denote high-efficiency performance (FoM ≈ 0.67). Conversely, dark blue-violet regions indicate configurations with poor fidelity or those falling within the exclusion zone where the FoM is set to zero. This latter condition is enforced when the absolute difference between the two focal lengths is less than 1500 μm (| *f(+H) - f(-H)* | ≤ 1500 μm), ensuring sufficient longitudinal separation between the focal planes to avoid cross-talk.

Outside this exclusion zone, the landscape reveals distinct high-intensity ridges radiating from the origin. These ridges correspond to combinations of *f(+H)* and *f(-H)* where the phase and amplitude requirements constructively align with the discrete phase and reflectance values available in our BIG nanopillar library. Along these ridges the metasurface is able to satisfy the two hyperboloidal phase profiles (Methods, Eq. 2) for both magnetic field directions, simultaneously with minimal phase and amplitude error.

The red star in Figure 2 identifies the best configurations for each period, for device design.

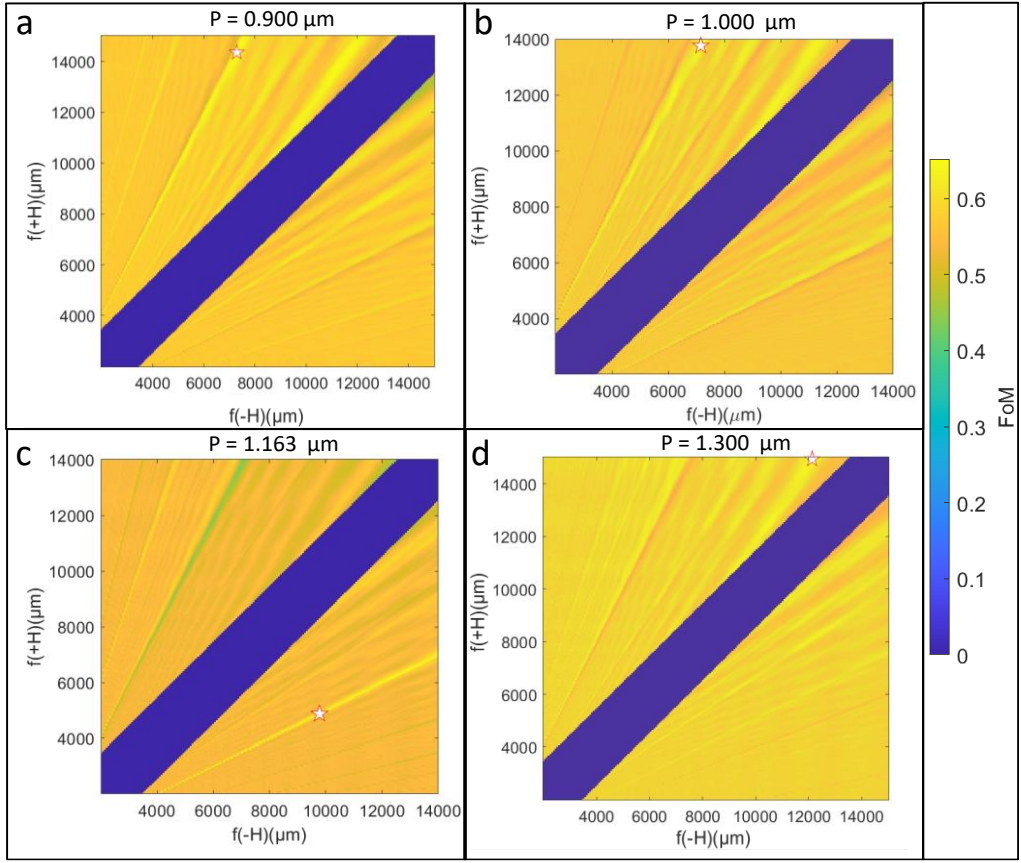

**Figure 2: Optimization landscapes for the bifocal metasurface design.** The 2D maps (a-d) displays the Figure of Merit (FoM) as a function of the target focal lengths *f(+H)* and *f(-H)* for each selected period of the unit cell. The colour scale indicates the harmonic mean of the focusing efficiencies, with bright yellow regions representing optimal performance. The dark diagonal band indicates the excluded region where the focal length difference is insufficient for distinct switching ($|\Delta f| \leq 1500$ μm). The red stars indicate the best design configurations.

## 2.3 Device Design and Gires-Tournois Amplification

Guided by this optimization landscape, we selected the configuration that offered the most promising balance of efficiency and focal separation for our case study. Focusing on the map for *P* = 1.000 μm (Fig.2 panel b), the optimization identified a good efficiency at the coordinates *f(+H)* = 7160.8 μm and *f(-H)* = 13758.8 μm. Consequently, we adopted this specific period and focal length couple as the reference design for the rigorous analysis presented in the following sections.

Applying the optimization framework described in Methods, we realized a final metasurface design capable of bi-focal operation at λ = 1.550 μm. The optimized structure consists of BIG cylindrical resonators arranged on a 0.2644 μm $SiO_2$ spacer, backed by an optically thick Au mirror.

This specific multilayer stack functions as a Gires-Tournois (GT) interferometer, which is critical for the device magnetic tunability[4]. Unlike single-pass transmissive metasurfaces[19], the GT configuration resonantly traps incident RCP light within the cavity. The resulting multiple internal reflections facilitate a multi-pass amplification of the magneto-optical effect, yielding non-reciprocal phase shifts significantly larger than those achievable in bulk material or single-layer metasurfaces. By spatially varying the BIG pillar diameters across the lens aperture, we effectively modulate this enhanced phase accumulation to satisfy the hyperbolic phase requirements for the two magnetic states simultaneously. Figure 3 shows the pillars spatial map and pillars diameters distribution of the designed metalens.

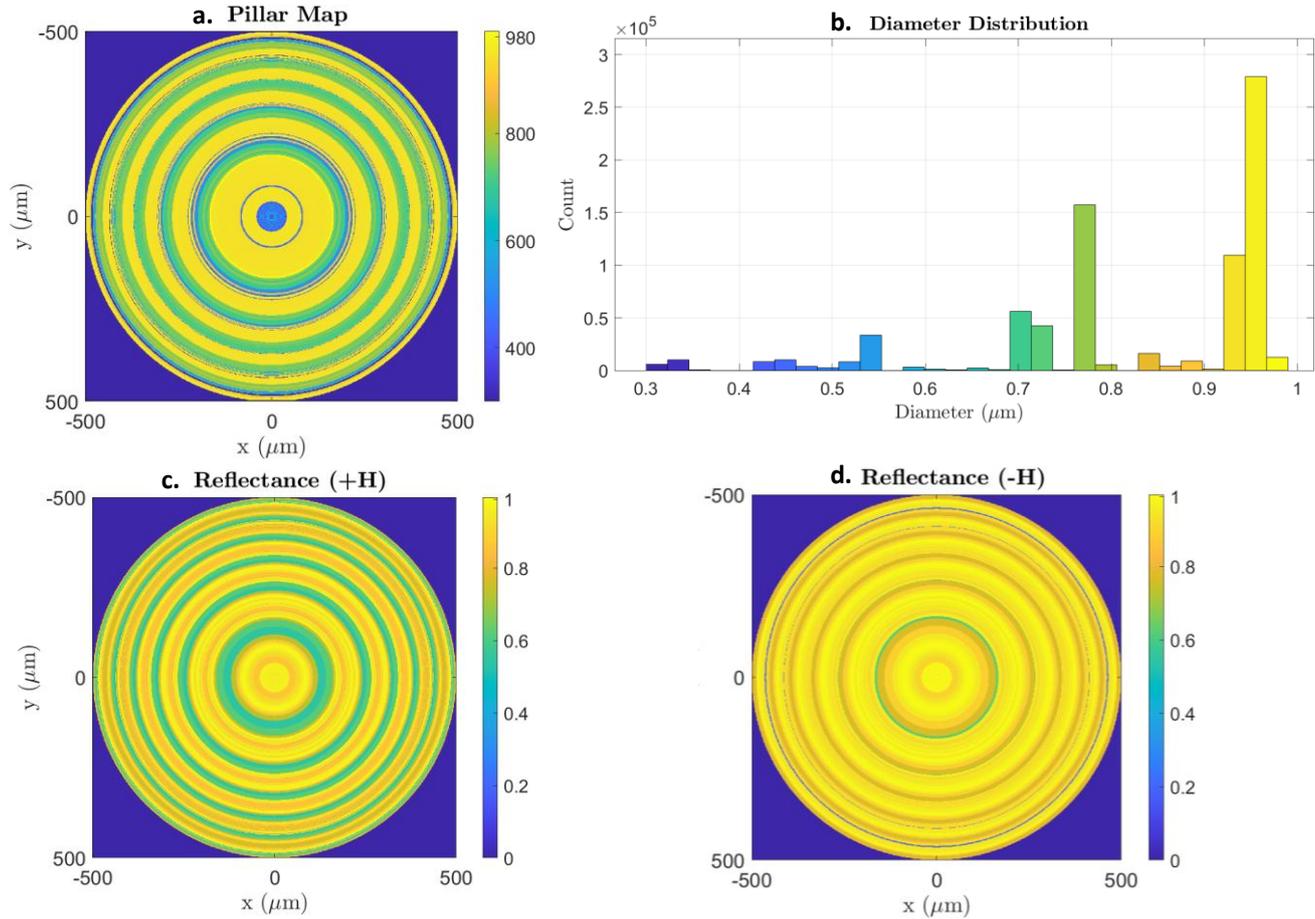

**Figure 3: Spatial pillar distribution and reflectance maps of the metasurface.** The plot shows the spatial distribution of the chosen pillar diameters across the 500 μm radius metasurface aperture (a-b). The concentric ring structure is characteristic of a Fresnel-type lens (a). The histogram shows the distribution of the implemented diameters (b). Spatial reflectance distribution of the metasurface under RCP illumination (c-d). The 2D maps illustrate the reflectance response across the x-y plane in the presence of (c) a positive external magnetic field (+$H$) and (d) a negative external magnetic field (–$H$).

A statistical analysis of the optimized geometry reveals a significant practical advantage: the algorithm converged on a solution that does not require a continuous range of diameters. Instead, the pillar dimensions are heavily clustered into specific bands, most notably around 0.700 - 0.800 μm and 0.900 - 1.000 μm. Given the fixed pillar height of 1.860 μm, these diameter ranges correspond to relatively low aspect ratios (AR) of approximately 2.5 to 2. This geometric clustering is highly advantageous for two reasons:

- Fabrication Feasibility: the lower aspect ratios significantly reduce the complexity of the manufacturing process, making the device compatible with standard high-resolution lithography and etching techniques.

- Simulation Accuracy: lower AR structures generally exhibit better agreement with the LPA used in our design phase, suggesting a high probability that experimental performance will closely match numerical predictions.

**2.4   Optical Performance and Focal Switching**

Figure 4 presents the reconstructed phase profiles at the metasurface plane (z = 0). The profiles clearly demonstrate that reversing the magnetic bias fundamentally alters the wavefront curvature. Specifically, the realized phase distributions correspond to two distinct focal lengths: *f(+H)* = 7160.8 μm and *f(-H)* = 13758.8 μm.

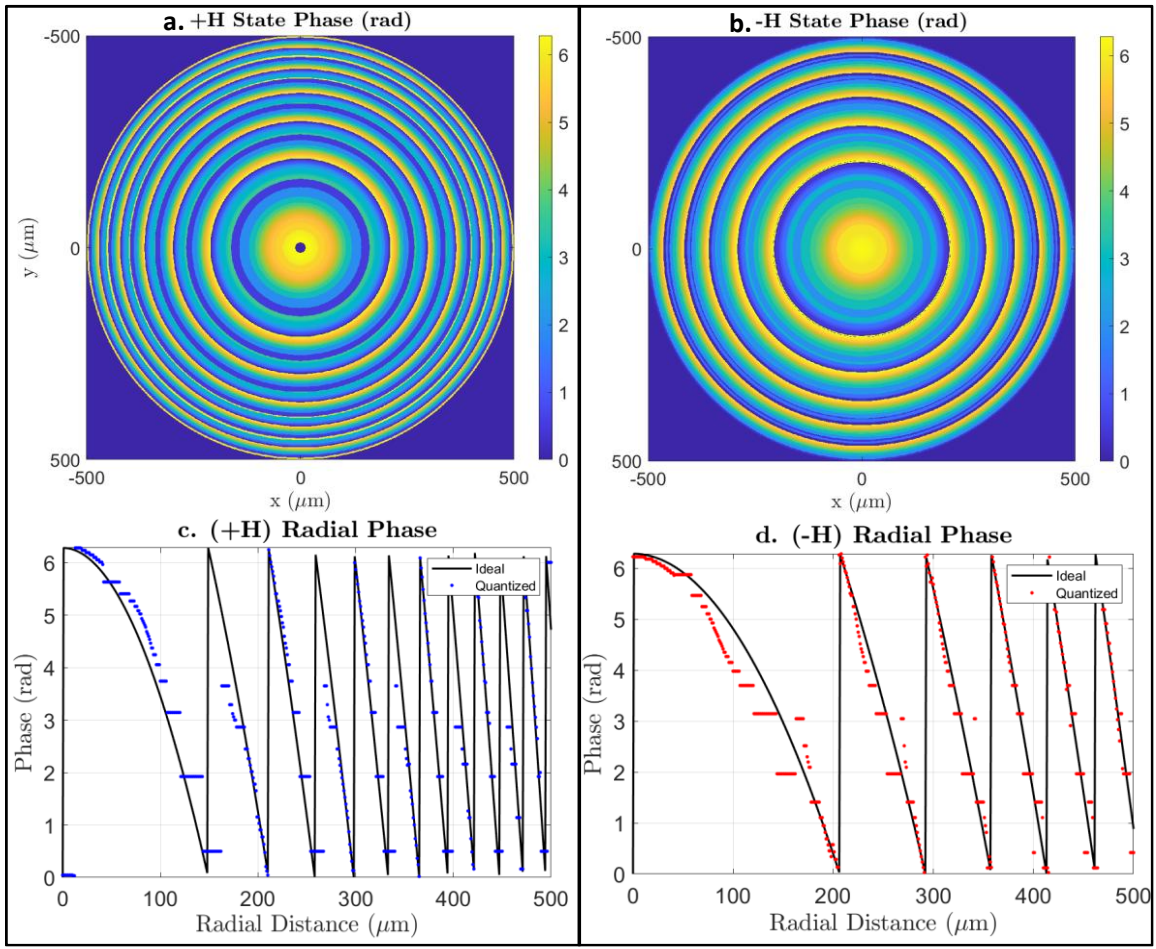

**Figure 4: Calculated phase profiles of the designed BIG-based bi-focal metalens.** Calculated

phase profiles of the designed BIG-based bi-focal metalens for H+ (a) and H- (b). The realized phase distributions correspond to focal lengths of: $f(+H)$ = 7160.8 μm and $f(-H)$ = 13758.8 μm.

The macroscopic performance of the designed metasurface was validated via the Angular Spectrum Method. The results confirm that magnetically switching between the two phase profiles (+$H$ and -$H$) successfully induces a precise longitudinal shift of a well-defined focal spot. As illustrated in the axial intensity distributions (Fig. 5c, 5f), the +$H$ state generates a sharp focus at the first target position with a focusing efficiency of 57%. Conversely, switching to the -$H$ state shifts the focus to the second target position, achieving a higher efficiency of 76%. Crucially, the simulation demonstrates high-contrast switching with minimal crosstalk at the inactive focal positions, achieved entirely without mechanical movement.

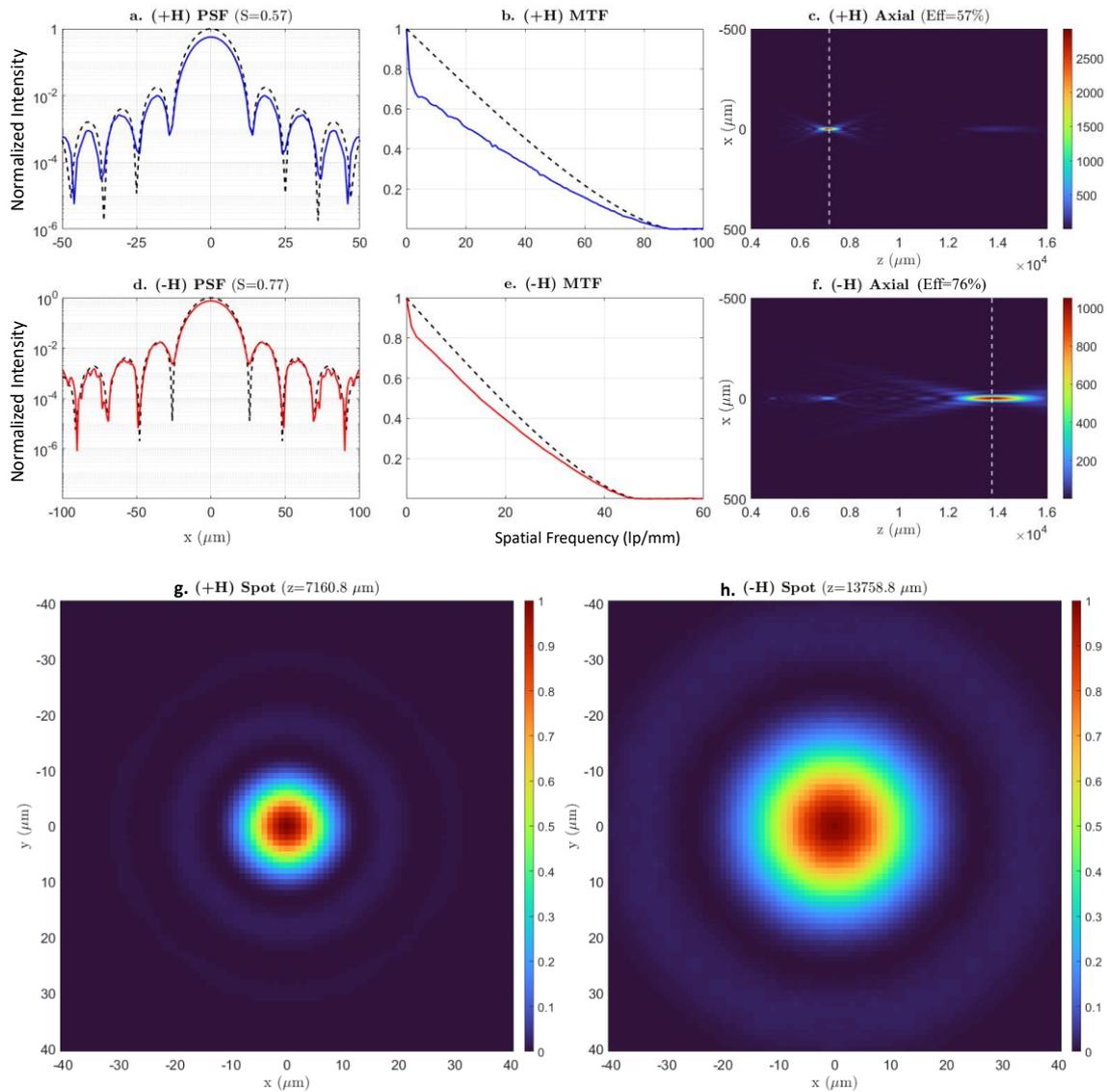

**Figure 5: Optical performance analysis of the bifocal metalens under magnetic switching.** (a–c) +H State: The lens targets the near focal length $f(+H)$. (a) Normalized Point Spread Function (PSF)

showing a Strehl ratio of 0.57 (blue solid) compared to the diffraction limit ideal case (black dashed). (b) Modulation Transfer Function (MTF). (c) Axial intensity distribution map (*x-z* plane) showing a clean primary focus with 57% efficiency. (d–f) -H State**:** The lens switches to the far focal length, $f(-H)$. (d) PSF showing a Strehl ratio of 0.77 (red solid). (e) MTF profile. (f) Axial intensity map confirming the longitudinal focal shift to the second target position with 76% efficiency. (g) Focus at z = 7160.8 µm, the spot is smaller, consistent with the shorter focal length. (h) Focus at z = 13758.8 µm, the lens produces a larger spot.

Figure 5 (g-h) reports the behaviour of the focal spot for the two different external magnetic field:
- +H (z = 7160.8 µm**):** The lens exhibits a tighter focus consistent with a shorter focal length, characterized by a Numerical Aperture (NA) of 0.070 and a focusing efficiency of 57%.
- -H (z = 13758.8 µm): In the longer focal length state, the lens produces a larger spot size with an NA of 0.036 and a focusing efficiency of 76%

## 2.5 Implications for Active Nanophotonics and Polarimetry

These results indicate that while the efficiency is primarily limited by material absorption and phase discretization, the device successfully provides distinct, field-tunable focusing capabilities suitable for highly integrated optical systems.

From reported results we outline that the inherent capacity of the proposed system to spatially decompose orthogonal spin states along the longitudinal axis, focusing RCP and LCP at distinct focal planes, addresses a fundamental challenge in integrated nanophotonics: the miniaturization of spin-sorting and polarimetric architectures. Because linearly polarized light is a coherent superposition of equal parts RCP and LCP, the proposed structure functions natively as a longitudinal spin beam splitter. Rather than relying on bulky, cascaded waveplates and polarizing beam splitters, the metasurface dynamically de-multiplexes the spin angular momentum of the incident field. Operating within the telecommunications C-band, this longitudinal separation is of critical practical importance for optical routing and polarization-division multiplexing. By isolating RCP and LCP components at different propagation depths, the device enables direct, crosstalk-free coupling into vertically stacked waveguides or multi-layer 3D photonic integrated circuits.

Furthermore, this spin-selective focusing provides a direct mechanism for compact, single-shot polarimetry. Measuring the differential focal intensity at the two distinct focal spots instantly quantifies the circular polarization content (the Stokes parameter) of an unknown incident beam without requiring temporal scanning. Alternatively, our systems can be exploited to carry out time multiplexed polarimetric imaging at frame rates not achievable by standard mechanical approach. As highlighted by recent state-of-the-art literature, exploiting these magneto-optical principles is critical for advanced light control in dual polarizations (Ignatyeva et al.[11]). The recent development of active magneto-optical metalenses further emphasizes that eliminating mechanical moving parts is essential for modern applications (Habibighahfarokhi et al.[19]). Ultimately, combined with the non-reciprocal phase accumulation inherent to the device, this bi-focal platform provides a highly robust, static framework for the realization of ultra-compact optical isolators, spin-selective optical interconnects, and dynamic routing architectures in next-generation photonic networks.

In comparison to existing literature, the proposed GT metalens architecture demonstrates a powerful strategy to enhance the intrinsic magneto-optical activity of BIG by exploiting multi-pass light-matter interactions within a resonant cavity. This configuration effectively transforms the inherently weak, single-pass MO phase shift into a significant and controllable modulation, enabling the realization of a non-mechanical bi-focal metalens with polarization- and field-dependent focusing characteristics. Indeed, these simulative results represent a distinct macroscopic advance over recent active magneto-optical metalenses. For instance, the transmissive, single-pass design recently proposed by Habibighahfarokhi et al.[19] yields a focusing efficiency of approximately 45% relative to the incident power, and primarily achieves an intensity modulation of about 11% at a fixed focal plane. Our architecture exploits GT reflective structures that amplify the MO effect beyond what is achievable in transmissive geometries. Consequently, our device surpasses previous performance metrics by delivering higher overall efficiencies (up to 76%) and achieving a massive longitudinal focal shift of a factor of roughly two, rather than modulating intensity at a single focus.

The numerical simulations confirm that the focal length of the designed BIG-based bifocal metalens can be magnetically tuned between $f(+H) = 7160.8$ μm and $f(-H) = 13758.8$ μm by reversing a modest saturation magnetic field of ±0.2 T, or alternatively, by switching the handedness of the incident circularly polarized light. This switching behaviour occurs without any mechanical adjustment, highlighting the strong MO modulation efficiency achieved through the GT enhancement. The corresponding numerical apertures of $NA_{H-} \approx 0.04$ with focusing efficiency of 76% and $NA_{H+} \approx 0.07$ with focusing efficiency of 57% indicate distinct focusing capabilities for the two polarization states. While the overall focusing efficiency remains moderate, primarily limited by absorption losses and imperfect phase discretization, the observed performance is already well within the range required for practical integrated and miniaturized optical systems. In particular, this level of efficiency is acceptable for polarization-selective beam routing, non-reciprocal optical elements, reconfigurable imaging modules, and on-chip optical interconnects, where dynamic tunability, compactness, and non-mechanical control outweigh the demand for absolute throughput.

Finally, we underline that the design strategy that we devised provides a fairly large flexibility in finding pairs of largely different focal lengths, with high focusing efficiency. Among the various possibilities, we explored in depth the features of a specific solution, but other combinations of focal lengths would perform about equally well. This flexibility is a significant advantage for the design of devices tailored for specific real-world applications.

These findings establish the GT metasurface platform as a promising route toward actively tunable magneto-optical metasystems, capable of operating with sub-millimeter precision under low intensity magnetic bias. The demonstrated capability to reversibly control focal length through either magnetic or polarization modulation represents a significant advance in the development of dynamic flat optics. This approach paves the way for future integration of non-reciprocal functionalities, MO-based adaptive lenses, and field-controlled photonic components, marking an important step toward multi-functional, magnetically reconfigurable metasurface technologies for next-generation photonic and optoelectronic applications.

## 3. Methods

For the design of the magnetically tunable metalens, we selected Bismuth Iron Garnet (BIG)[20–22] as the active medium. Under a saturation magnetization field of H = 0.2 T[20] applied along the z-axis, the material permittivity tensor is given by:

$$\hat{\varepsilon}(\mathbf{B}) = \begin{pmatrix} \varepsilon_{xx}(\mathbf{B}) & \varepsilon_{xy}(\mathbf{B}) & 0 \\ -\varepsilon_{xy}(\mathbf{B}) & \varepsilon_{yy}(\mathbf{B}) & 0 \\ 0 & 0 & \varepsilon_{zz}(\mathbf{B}) \end{pmatrix} \quad (1)$$

As we can see from Eq.(1), the dielectric response of magnetized BIG at 1550 nm is characterized by a tensor with diagonal term $\varepsilon_{xx} = \varepsilon_{yy} = \varepsilon_{zz} = 6.25$ and off-diagonal term $\varepsilon_{xy}(\mathbf{B}) = i \cdot \mathbf{g}$ where $\mathbf{g}$ is the gyrotropic constant equal to $0.06$[18].

The off-diagonal terms break time-reversal symmetry, enabling magneto-optical modulation of the cavity mode.

### 3.1 Unit Cell Characterization

Numerical simulations were performed in COMSOL Multiphysics® using the Finite Element Method[23] (FEM). The unit cell consists of a BIG cylindrical nanoresonator (with height, $h_{cyl}$ = 1860 nm) placed on a silica spacer backed by an Au reflector interacting with normally incident RCP light. The thickness of the silica spacer is set to $t_d = \frac{\lambda}{4n_{BIG}} = 264.4$ nm.

An infinite periodic array was modelled using periodic boundary conditions with a fixed lattice period of $P = 1000$ nm. To characterize the optical response as a function of geometry, a parametric analysis was performed over the nanoresonator diameter $D$, which was varied from 300 nm to 980 nm in increments of 2 nm. In this way a series of distinct infinite periodic arrays, each defined by a unique nanoresonator diameter, was analyzed.

For each value of $D$, the complex reflection coefficient of the nanopillar arrays, including both amplitude and phase, was calculated for two magnetic configurations corresponding to positive ($H = +0.2$ T) and negative ($H = -0.2$T) magnetic saturation. The resulting data were collected into a lookup table linking the nanoresonator diameter to its corresponding optical response in both magnetic states. This library served as the basis for the subsequent metasurface design using the Local Periodic Approximation (LPA)[24].

## 3.2   Multi-State Metalens Optimization

The primary objective of the numerical simulations was to design a metasurface with a fixed geometry that can achieve two distinct focal lengths, *f(+H)* and *f(-H)*, depending on whether an external magnetic field of *+H* or *-H* is applied, under illumination by normally incident RCP light. For an ideal metalens focusing a planar wavefront at a target focal length $f_{target}$, the required phase profile $\varphi_{ideal}$ at a radial *r* from the optical axis is governed by the hyperboloidal phase profile:

$$\varphi_{ideal}(r) = \left(\frac{2\pi}{\lambda}\right)\left[\sqrt{r^2 + f_{target}^2} - f_{target}\right] mod(2\pi) \qquad (2)$$

Since the physical distribution of pillar diameters, *D(r)*, is fixed, it must simultaneously satisfy the phase requirements for both magnetic states. To address this, we implemented a dual-layer optimization algorithm.

Rather than arbitrarily selecting target focal lengths, we performed a parametric sweep to identify the optimal pair (*f(+H)*, and *f(-H)*) that maximizes the overall device performance. We defined a Figure of Merit (FoM) based on the harmonic mean of the focusing efficiencies in both states:

$$FoM(f(+H), f(-H)) = \frac{2\eta(+H)\cdot\eta(-H)}{\eta(+H)+\eta(-H)} \qquad (3)$$

where η represents the focusing efficiency, calculated as the ratio of the peak intensity at the focal point generated by our designed metasurface to the peak intensity that would be generated by an ideal aberration-free lens (with 100% reflectance and perfect continuous phase) of the same aperture (i.e. the Strehl ratio[25]). This metric ensures that the device performs consistently well in both states, penalizing pairs where one state is significantly less efficient than the other.

To accelerate the optimization, we exploited the rotational and reflective symmetry of the circular aperture. Calculations were performed over a single octant ($0 \leq \theta \leq \pi/4$), representing $1/8^{th}$ of the total metasurface area.

For a given pair of focal lengths, the physical geometry at each radial coordinate $r$ was determined by minimizing a composite cost function representing the Euclidean distance in the complex plane between the local ideal phase shift and the actual phase shift introduced by the nanopillar. For every candidate diameter $D$ in our library, the cost $C(D)$ was calculated as:

$$C(D) = \left|e^{i\varphi_t(r,+H)} - A e^{i\varphi_s(D,+H)}\right|^2 + \left|e^{i\varphi_t(r,-H)} - A e^{i\varphi_s(D,-H)}\right|^2 \tag{4}$$

Here, $A = \sqrt{R(D, \pm H)}$ is the amplitude of the reflected field under external magnetic fields of $\pm H$, while $\varphi_t(r, \pm H)$ denotes the target phase for these respective fields. The term $\varphi_s(D, \pm H)$ corresponds to the phase response extracted from COMSOL Multiphysics® simulations. This approach prioritizes phase matching while simultaneously suppressing choices that would result in low optical efficiency.

### 3.3  Numerical Validation via Angular Spectrum Method

To validate the focusing performance of the optimized design, we employed the Angular Spectrum Method (ASM)[26]. This wave propagation technique allows for the efficient analysis of large-scale metasurfaces that are computationally prohibitive for full-wave solvers.

First, the near-field distribution $E_{lens}(x,y)$ at the metasurface plane ($z = 0$) was reconstructed using the amplitude A ($\sqrt{R(D)}$) and phase $\varphi$ values retrieved from the lookup table for the selected diameters:

$$E_{lens}(x, y) = A e^{i\varphi_{sim}} \tag{5}$$

To simulate the focusing behavior, the field was propagated to the observation domain using the Fourier optics principle. The field $E(x,y,z)$ at any distance $z$ is computed by:

$$E_{lens}(x, y, z) = \mathcal{F}^{-1}\{\mathcal{F}[E_{lens}] \cdot H(k_x, k_y; z)\} \tag{6}$$

where $\mathcal{F}$ and $\mathcal{F}^{-1}$ denote the Fourier transform and its inverse[27], respectively. The free-space transfer function $H$ is defined as:

$$H(k_x, k_y, k_z) = e^{iz\sqrt{k_0^2 - k_t^2}} \tag{7}$$

A critical challenge in numerical propagation is the handling of high-spatial-frequency components. As the transverse wave vector magnitude $k_t = \sqrt{k_x^2 + k_y^2}$ approaches the free-space wavenumber $k_0 = \frac{2\pi}{\lambda}$, the transfer function oscillates rapidly. Furthermore, components where $k_t > k_0$ represent evanescent waves which decay exponentially rather than propagate.

To ensure numerical stability and physical accuracy, our implementation incorporates a spectral band-limiting filter directly within the transfer function calculation[28]. We enforce a hard cut-off condition for the transfer function $H$:

$$H(k_x, k_y, k_z) = \begin{cases} exp\left(iz\sqrt{k_0^2 - k_t^2}\right) & if \ k_t \leq k_0 \\ 0 & if \ k_t > k_0 \end{cases} \tag{8}$$

This operation serves two distinct purposes:

1. Evanescent Wave Filtering: by strictly setting the transfer function to zero for $k_t > k_0$, we explicitly remove non-propagating evanescent orders. This prevents the numerical instability that can arise from calculating the square root of negative values and eliminates high-frequency noise that does not contribute to the far-field focus.

2. Band-Limited Propagation: This acts as an ideal low-pass filter in the frequency domain, effectively band-limiting the signal to the propagating spectrum. This ensures that the reconstruction of the field at the focal plane is free from aliasing artefacts that would otherwise arise from the discretized sampling of sharp discontinuities at the pillar edges.

In Figure 6 we report a schematic representation of the used methodology for the design of the bifocal metalens:

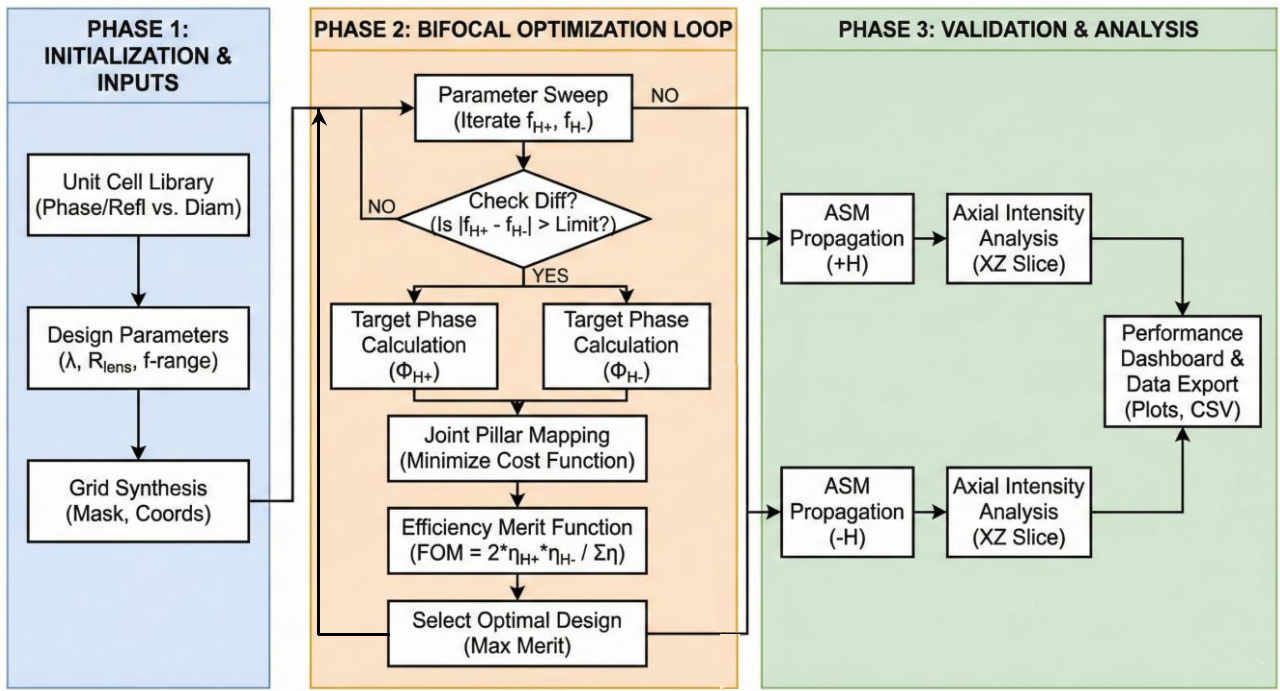

**Figure 6: Schematic workflow of the bifocal metasurface design and the analysis framework.** The process is divided into three phases: (1) Initialization, where unit cell libraries for the two different magnetic field orientations (*+H, -H*) are imported alongside lens geometry constraints; (2) Optimization Core, which employs a parameter sweep over focal lengths (*f(+H)* and *f(-H)*) to minimize a joint cost function involving phase error and efficiency penalties for both magnetic fields and (3) Validation, utilizing the Angular Spectrum Method (ASM) to propagate fields and generate axial intensity profiles to verify the distinct focal shifts for the optimized design.

**Data availability**

All simulated data that were used to produce the results reported in this Article are available via request to the Corresponding Authors.

**Code availability**

The custom MATLAB algorithms, the scripts and COMSOL Multiphysics models used to generate, optimize, and evaluate the metasurface designs in this study are available via request to the Corresponding Authors.

**Author contributions**

All authors have accepted responsibility for the entire content of this manuscript, consented to its submission, reviewed all the results, and approved the final version. A.S. and G.T. jointly developed the methodological approach. A.S. wrote the original draft, prepared the figures, and performed the simulations. B.P., A.G., and F.P. contributed ideas, discussed the results, and co-wrote the manuscript.

**Funding**

This activity has been supported by:
- the Regione Toscana grant under the program "Progetti di Alta Formazione attraverso l'attivazione di assegni di ricerca", European Social Fund 2021–2027. Link: https://www.regione.toscana.it/-/transizione-verde-finanziamenti-per-progetti-di-alta-formazione-con-assegni-di-ricerca.
- The European Union under the European Innovation Council grant agreement N. 101161583, project GreenSWap, link https://greenswapspace.eu/.

**Competing interests**

The authors declare no competing interests

**Correspondence and requests for materials** should be addressed to Corresponding Authors.